\documentclass[showpacs,preprintnumbers]{revtex4}
\usepackage{graphicx}
\usepackage{amssymb}
\usepackage{amsmath}

\input{tcilatex}

\begin{document}

\title{Pseudogap Phase: Exchange Energy Driven vs. Kinetic Energy Driven}
\author{Zhengcheng Gu, Tao Li, and Zheng-Yu Weng}
\affiliation{Center for Advanced Study, Tsinghua University, Beijing 100084, China}

\begin{abstract}
We show that \emph{both} kinetic and superexchange energies of the $t-J$
model may be read off from the optical data, based on an optical sum rule
for the Hubbard model. Then we comparatively study two mean-field theories
of pseudogap phase based on the $t-J$ model. We find that while the
pseudogap phase is superexchange-energy-driven in the slave-boson
resonating-valence-bond (RVB)\textbf{\ }state, it is kinetic-energy-driven
in the bosonic RVB state. The sharp contrast in the mechanisms of the
pseudogap phases can be attributed to the fact that the antiferromagnetic
(AF) correlations behave quite differently in two mean-field states, which
in turn distinctly influence the kinetic energy of charge carriers. We
elaborate this based on some detailed studies of the superexchange energy,
kinetic energy, uniform spin susceptibility, equal-time spin correlations,
dynamic spin susceptibility, as well as the optical conductivity. The
results provide a consistent picture and understanding on two physically
opposite origins of pseudogap phase. In comparison with experimental
measurements, we are led to conclude that the pseudogap phase in the
cuprates should be kinetic-energy-driven in nature.
\end{abstract}

\maketitle

%\section*{\protect\bigskip }

\section{\protect\bigskip Introduction}

The pseudogap phase is one of the most interesting and unconventional
regimes observed in the underdoped cuprate superconductors. Such a
phenomenon has been identified \cite{timusk} in NMR, ARPES,
neutron-scattering, transport, optical conductivity and other experiments,
which are absent in the traditional BCS superconductivity. The physical
origin of such a pseudogap state remains controversial, with the proposed
ones ranging from the RVB pairing \cite{f-RVB1,f-RVB2,f-RVB3}, the loss of
the phase coherence in the superconducting order parameter \cite{lps1,lps2},
preformed electron pairing \cite{preform}, to the d-density-wave theory \cite%
{ddw}, etc. A good understanding of the essential physics involved in the
pseudogap phase will be of great importance in searching for a sensible
microscopic theory of the high-$T_{c}$ cuprates.

In this paper, we propose to study the detailed driving mechanisms behind
the pseudogap phase. So far, different driving mechanisms for the
superconducting condensation have been studied extensively in literature 
\cite{super1,super2,super3,super4,super5,super6,super7,so51,so52}. As to be
shown in this work, the nature of a pseudogap phase can be also meaningfully
clarified by identifying, both experimentally and theoretically, whether it
is kinetic-energy-driven or potential-energy-driven. Specifically we shall
focus ourselves on the pseudogap phases based on the $t-J$ model. Such a
model is composed of two terms: $H_{t-J}=H_{t}+H_{J}$ (see Sec. II), where
the kinetic term is denoted by $H_{t}$ and the superexchange term by $H_{J}$%
. Then one can ask whether the pseudogap state, if exists, is mainly driven
by $H_{t}$ or $H_{J},$ and how different possible driving mechanisms can be
probed and verified (falsified) by experimental measurements.

Firstly we shall establish a very general relationship between the optical
conductivity and the kinetic and superexchange energies in the $t-J$ model.
An important but subtle difference between the kinetic terms of the $t-J$
model and the Hubbard model will be carefully examined and distinguished. In
a single-band Hubbard model, the optical sum rule is generally given by\cite%
{sumrule} 
\begin{equation}
\int_{0}^{\Lambda }\sigma _{1}\left( \omega ,T\right) d\omega =\dfrac{\pi
a^{2}}{2V}\left\langle -H_{\mathrm{K}}\right\rangle  \label{sumrule1}
\end{equation}%
where $\sigma _{1}\left( \omega ,T\right) $ is the real part of the optical
conductivity and $\Lambda $ is some high-energy cutoff (higher than the
Hubbard-Mott gap). On the right-hand-side (rhs), $H_{\mathrm{K}}$ is the
kinetic energy of the Hubbard model, $a$ is the lattice constant, and $V$ is
the volume. In the case of the $t-J$ model (\emph{i.e.,} the large-$U$
Hubbard model), the kinetic energy $\left\langle H_{\mathrm{K}}\right\rangle 
$ can be replaced by $\left\langle H_{t}\right\rangle ,$ if $\Lambda $ is
taken as some characteristic energy cutoff $\Omega $ below which \emph{only}
no-double-occupancy intraband transitions of charges are allowed. However,
if $\Lambda $ is taken to be larger such that to allow the transitions to
doubly occupied states, then the rhs of Eq.(\ref{sumrule1}) will include
additional contribution beside $\left\langle H_{t}\right\rangle $ from the
kinetic energy of the Hubbard model, which, as we will show in Sec. II, is
nothing but $2\left\langle H_{J}\right\rangle ,$ twice of the superexchange
energy of the $t-J$ model!

Thus, by properly choosing the magnitudes of $\Lambda $ and $\Omega ,$
respectively, the optical data can provide crucial information on \emph{both}
the kinetic energy and superexchange energy of the $t-J$ model and be
effectively used to deduce the driving mechanisms for both pseudogap and
superconducting states. As to be discussed in Sec. II, the recent optical
measurement results \cite{exp} in the cuprate $Bi_{2}Sr_{2}CaCu_{2}O_{8}$
can be read off, based on the present analysis, as strong evidence in
support of the kinetic-energy-driven mechanism for \emph{both}
superconducting\emph{\ }and pseudogap phases.

Then we shall study two distinct pseudogap phases defined in different
mean-field theories of the $t-J$ model, \emph{i.e.,} the slave-boson RVB 
\cite{sb1,sb2} and bosonic RVB (b-RVB) \cite{string1,string2} mean-field
states, respectively, which will be introduced in Sec. III. Then, we show
that the pseudogap phase in the slave-boson RVB state is
superexchange-energy-driven, while it is kinetic-energy-driven in the b-RVB
state. The underlying physics is then explored in a comparative way in Sec.
IV.

A detailed study of the static and dynamic spin susceptibility functions in
Sec. IV reveals how the AF correlations evolve between the pseudogap and
\textquotedblleft normal\textquotedblright\ phases. In the slave-boson RVB
state, the AF correlations near $(\pi ,\pi )$ are intrinsically weak in the
\textquotedblleft normal\textquotedblright\ state and by opening the RVB gap
to enter the pseudogap phase, the superexchange energy will be reduced,
whereas the kinetic energy simultaneously gets increased. In contrast, the
AF correlations near $(\pi ,\pi )$ are already quite strong in the
\textquotedblleft normal\textquotedblright\ state of the b-RVB state such
that the kinetic energy is very frustrated. By going into the pseudogap
phase, the kinetic energy will get improved with suppressing AF
correlations, which simultaneously results in an increase of the
superexchange energy. Thus, whether a pseudogap phase in the $t-J$ model is
kinetic-energy-driven or superexchange-energy-driven crucially depends on
which one of the kinetic energy and AF correlations gets \textquotedblleft
released\textquotedblright\ from a \textquotedblleft
suppression\textquotedblright\ status in the \textquotedblleft
normal\textquotedblright\ state.

Experimentally, the superexchange energy of$\ H_{J}$ will be related to the
nearest-neighbor (NN) spin-spin correlations, which can be decided by the
imaginary dynamic spin susceptibility $\func{Im}\chi \left( \overrightarrow{q%
},\omega \right) ,$ measurable by an inelastic neutron-scattering
measurement, through the following sum rule 
\begin{equation}
\left\langle \vec{S}_{i}\cdot \vec{S}_{j}\right\rangle =\int \frac{d^{2}\vec{%
q}}{\left( 2\pi \right) ^{2}}\int_{0}^{\infty }\frac{d\omega }{\pi }%
(1+n(\omega ))\func{Im}\chi \left( \vec{q},\omega \right) \cos \left( \vec{q}%
\cdot \left( \vec{r}_{i}-\vec{r}_{j}\right) \right)  \label{sumrule2}
\end{equation}%
where $n(\omega )=1/(e^{\beta \omega }-1).$ Previously, based on this
relation, arguments for superexchange-energy-driven superconductivity have
been made in literature based on the neutron-scattering measurements. In
particular, the SO(5) theory suggests \cite{so51,so52} that the
superconducting transition in the high-$T_{c}$ superconductors is driven by
the superexchange energy gain as the so called $\pi $- resonance mode \cite%
{neutron1} opens a new channel for AF spin fluctuations in the
superconducting state. But for both the slave-boson RVB and b-RVB theories
of the $t-J$ model, the superconductivity is always kinetic-energy-driven.
The b-RVB theory goes further to predict that the pseudogap phase is also
kinetic-energy-driven, in which the resonance-like peak observed in $\func{Im%
}\chi \left( \vec{q},\omega \right) $ by neutron-scattering \cite%
{neutron1,neutron2,so53} is not interpreted as an emergent new mode, rather
it simply comes from the low-lying spin excitations in the \textquotedblleft
normal\textquotedblright\ state, which are pushed upwards to a \emph{higher
energy }and become sharpened in the pseudogap phase. In other words, if the
\textquotedblleft normal\textquotedblright\ state is \textquotedblleft
extrapolated\textquotedblright\ into zero temperature, strong AF
correlations near $\omega \sim 0$ or even an AF long-range order can appear
as evidenced by the non-Korringa behavior of the NMR $1/T_{1}T$ both
calculated (Sec. IV) and observed experimentally in the cuprates. Therefore,
in combination with the aforementioned optical measurements, the experiments
seem overall in favor of the mechanism of kinetic-energy-driven as the
origin of the pseudogap phase.

In Sec. II, the optical sum rules for the Hubbard and $t-J$ models are
discussed and a comparison with experiment is made. In Sec. III, we
introduce and discuss two pseudogap phases in the slave-boson RVB and b-RVB
theories based on the $t-J$ model, and in Sec. IV, we present a comparative
study of the pseudogap phases in two RVB mean-field states. Finally a
discussion and conclusions are given in Sec. V.

\section{Optical sum rules in the Hubbard and $t-J$ models}

The Hubbard model is defined by

\begin{equation*}
H_{\mathrm{Hub}}\equiv H_{\mathrm{K}}+H_{U}=-t\dsum\limits_{\left\langle
ij\right\rangle \sigma }(c_{i\sigma }^{\dagger }c_{j\sigma
}+h.c.)+U\dsum\limits_{i}n_{i\uparrow }n_{i\downarrow }.
\end{equation*}%
In the large-$U$ limit, the Hubbard model can be reduced \cite{bza} to the $%
t-J$ model in the low-energy subspace of no double occupancy. This is
usually done by dividing the Hilbert space into the low-energy subspace of
no double occupancy and the high-energy subspace with doubly occupied sites.
In the large-$U$ limit, two subspaces are separated by a gap of order of $U$%
. Correspondingly, the Hamiltonian can be divided into intra-subspace and
inter-subspace pieces. Introducing the projection operator $P_{L}$ and $P_{H%
\text{ }}$for the low-energy subspace and high-energy subspace,
respectively, the Hubbard Hamiltonian can be written as

\begin{equation*}
H_{\mathrm{Hub}}=H_{L}+H_{H}+H_{mix}
\end{equation*}%
in which

\begin{align*}
H_{L}& =P_{L}H_{\mathrm{K}}P_{L} \\
H_{H}& =P_{H}H_{\mathrm{K}}P_{H}+H_{U} \\
H_{mix}& =P_{L}H_{\mathrm{K}}P_{H}+P_{H}H_{\mathrm{U}}P_{L}
\end{align*}%
are the terms in the low-energy subspace, high-energy subspace, and the
subspace-mixing term, respectively. The subspace-mixing term can be removed
by a canonical transformation $e^{iS}$ \cite{bza}. To the first order of $%
\frac{t}{U}$, the transformed Hamiltonian in the low energy subspace is the
standard $t-J$ model

\begin{align*}
\left( e^{iS}H_{\mathrm{Hub}}e^{-iS}\right) _{L}& =P_{L}H_{\mathrm{K}%
}P_{L}+iP_{L}\left[ S,H_{mix}\right] P_{L} \\
& \equiv H_{t-J}
\end{align*}%
where $H_{t-J}=H_{t}+H_{J},$ with the kinetic term $H_{t}$ and the
superexchange term $H_{J}$ defined by

\begin{equation}
H_{t}=-t\dsum\limits_{\left\langle ij\right\rangle \sigma }\left( \hat{c}%
_{i\sigma }^{\dagger }\hat{c}_{j\sigma }+h.c.\right) ,
\end{equation}%
\begin{equation}
H_{J}=J\dsum\limits_{\left\langle ij\right\rangle }\left( \vec{S}_{i}\cdot 
\vec{S}_{j}-\frac{1}{4}n_{i}n_{j}\right) ,
\end{equation}%
where $\hat{c}_{i\sigma }=\left( 1-n_{i-\sigma }\right) c_{i\sigma }$ (here $%
n_{i\sigma }=c_{i\sigma }^{\dagger }c_{i\sigma }$), $\vec{S}_{i}=\frac{1}{2}%
\sum_{\alpha \beta }c_{i\alpha }^{\dagger }\left( \vec{\sigma}\right)
_{\alpha \beta }c_{i\beta },$ and $n_{i}=\sum_{\sigma }n_{i\sigma }$.

Under the above canonical transformation, the kinetic energy of the Hubbard
model is transformed into%
\begin{equation}
\left( e^{iS}H_{\mathrm{K}}e^{-iS}\right)
_{L}=H_{t}+2J\dsum\limits_{\left\langle i,j\right\rangle }\left( S_{i}\cdot
S_{j}-\frac{1}{4}n_{i}n_{j}\right)  \label{k}
\end{equation}%
in the low-energy subspace to the first order of $\frac{t}{U}$. The
potential energy of the Hubbard model transforms into

\begin{equation*}
\left( e^{iS}H_{U}e^{-iS}\right) _{L}=-J\dsum\limits_{\left\langle
i,j\right\rangle }\left( S_{i}\cdot S_{j}-\frac{1}{4}n_{i}n_{j}\right)
=-H_{J}
\end{equation*}%
in the low-energy subspace to the same order.\ Note that the transformed
form of $H_{U}$ is still positive-definite.

From these formulas, we can explicitly see an important but subtle
difference between the kinetic energy in the $t-J$ model and that of the
underlying Hubbard model. Note that $H_{t}$ solely contributes to the charge
response in the $t-J$ model with $H_{J}$ as charge neutral in the subspace
of no double occupancy. According to Eq.(\ref{k}), however, the $\emph{%
kinetic}$ \emph{energy} of the Hubbard model includes not only the kinetic
term $H_{t}$ of the $t-J$ model, but also an \emph{additional} term, which
happens to be twice of the superexchange term $H_{J}$. Namely, the
\textquotedblleft kinetic energy\textquotedblright\ term $H_{t}$ in the $t-J$
model does not represent the \emph{whole} kinetic energy of the Hubbard
model. Physically, this is due to the fact that no matter how large $U$ is,
as long as it is finite, the lower-Hubbard band does not exactly correspond
to the no double occupancy subspace and, to the leading order approximation
of $t/U$, the correction $2H_{J}$ in Eq.(\ref{k}) originates from the \emph{%
virtual} transitions of charge carriers to the doubly occupied states, which
only vanishes at $U=\infty $ or $J=0$. Therefore, the kinetic energy at the
level of the Hubbard model is different from the kinetic energy in the $t-J$
model. Due to such a subtle difference, one has to distinguish the
terminology of \textquotedblleft kinetic energy driven vs. superexchange
energy driven\textquotedblright\ at the level of the $t-J$ model from
\textquotedblleft kinetic energy driven vs. potential energy
driven\textquotedblright\ at the level of the Hubbard model to avoid
confusion, which has been correctly noted \cite{super6} before.

In the optical sum rule, the kinetic energy of the Hubbard model is related
to the total optical response in Eq.(\ref{sumrule1}). According to the
preceding discussion, the total optical weight of the Hubbard model is then
composed of two contributions, namely the optical response within the
subspace of no double occupancy and the optical response involving doubly
occupied sites.

The first contribution is just the optical response of the $t-J$ model,
contributed by $H_{t}.$ According to the optical sum rule, this contribution
is related to kinetic energy of the $t-J$ model in the following way

\begin{equation}
\dint\nolimits_{0}^{\Omega }\sigma _{1}\left( \omega \right) d\omega =\dfrac{%
\pi a^{2}}{2V}\left\langle -H_{t}\right\rangle .  \label{op2}
\end{equation}%
Here an energy cutoff $\Omega $ is a characteristic scale below which the
charge response of the Hubbard model is equivalent to that of the $t-J$
model.

The additional kinetic energy in Eq. (\ref{k}) involving doubly occupied
states should be determined by the optical weight between $\Omega $ and $%
\Lambda $ as given by

\begin{eqnarray}
\dint\nolimits_{\Omega }^{\Lambda }\sigma _{1}\left( \omega \right) d\omega
&=&\dfrac{\pi a^{2}}{2V}\left\langle -2J\dsum\limits_{\left\langle
i,j\right\rangle }\left( S_{i}\cdot S_{j}-\frac{1}{4}n_{i}n_{j}\right)
\right\rangle  \notag \\
&=&\dfrac{\pi a^{2}}{2V}\left\langle -2H_{J}\right\rangle  \label{op3}
\end{eqnarray}%
which is thus related to the superexchange energy of the $t-J$ model!
Therefore, if we can determine the energy cutoff $\Lambda $ and $\Omega $,
we can read off both the kinetic energy and the superexchange energy of the $%
t-J$ model from the optical data.

Recently, optical measurements in the cuprate $Bi_{2}Sr_{2}CaCu_{2}O_{8}$
have revealed \cite{exp} a continuous spectral weight transfer from the high
energy part to the low energy part of the in-plane optical conductivity with
decreasing temperature, in both optimally doped and underdoped samples. An
characteristic energy scale $10^{4}$ $\mathrm{cm}^{-1}$ has been identified,
which corresponds to the minimum of the in-plane optical conductivity in the
experiment. Below and above it, the integrated optical conductivity behaves
in opposite ways as a function of temperature. Namely, the spectral weight
between $10^{4}$ $\mathrm{cm}^{-1}$ to $2\times 10^{4}$ $\mathrm{cm}^{-1}$
transfers steadily to below $10^{4}$ $\mathrm{cm}^{-1}$ with decreasing
temperature, while the total spectral weight remains approximately conserved 
\cite{exp}.

If we take $10^{4}$ $\mathrm{cm}^{-1}$ as the energy cutoff $\Omega $ for
the $t-J$ model described above and adopt $2\times 10^{4}$ $\mathrm{cm}^{-1}$
as $\Lambda ,$ below which reliable optical data is available \cite{exp},
then the experimental result consistently indicates that, starting well
above and persisting below $T_{c},$ the kinetic energy of the $t-J$ model is
continuously lowered whereas the superexchange energy is increased with
decreasing temperature according to Eqs.(\ref{op2}) and (\ref{op3}). Such
experimental results can be thus read off as evidence in support of the
kinetic-energy-driven mechanism for \emph{both} superconducting\emph{\ }and
pseudogap phases, so long as the $t-J$ model is relevant to the cuprates
with the energy cutoff of $\Omega \sim 10^{4}$ $\mathrm{cm}^{-1}$
representing the characteristic energy scale for the intraband transitions
within the no-double-occupancy subspace. Note that the value of the energy
cutoff $\Lambda $ of the Hubbard model is subjected to some uncertainty, but
this may not matter much since\ the spectral weight approximately remains
conserved as a function of temperature with $\Lambda \gtrsim 2\times 10^{4}$ 
$\mathrm{cm}^{-1}$ as stated above.

The steady increase of the superexchange energy with decreasing temperature
seems at odd with the conventional understanding of the pseudogap as some
kind of spin pairing gap forming, driven by the exchange energy of the $t-J$
model. This indicates that the kinetic energy of the $t-J$ model plays an
important role in the formation of the pseudogap \cite{string4,pseudo2}. In
the next section, we shall investigate the detailed driving mechanisms for
pseudogap phases in the $t-J$ model based on two different mean-field
theories and explore the underlying physics in comparison with experiment.

Finally we note that a closer examination on the experimental data shows
that both kinetic and superexchange energies change more rapidly upon
entering the superconducting state. This shows convincingly that the
superconducting transition is kinetic energy driven while the exchange
energy resists the transition. Also, we note that since the kinetic energy
and superexchange energy evolve in opposite directions with temperature, the
true superconducting condensation energy should be smaller than that
estimated from the kinetic energy alone. Furthermore as the superconducting
transition usually simultaneously involves the interlayer coherent hopping
in the cuprates, one needs to further consider the $c$-axis conductivity in
order to determine whether the intra or interlayer kinetic energy will
predominantly ``drive'' the superconductivity\cite{c-axis1,c-axis2}, if it
is kinetic-energy-driven. Nevertheless, in the present work we shall be
mainly interested in the formation of the pseudogap phase above $T_{c}$,
where the out-of-plane physics is no longer as crucial, and one can thus
mainly focus on the in-plane physics.

\section{Two distinct pseudogap phases based on the $t-J$ model}

\subsection{Pseudogap phase in slave-boson RVB state}

In the slave-boson formalism of the $t-J$ model, the electron operator can
be written as \cite{sb1} 
\begin{equation}
c_{i\sigma }=h_{i}^{\dagger }f_{i\sigma }  \label{sb1}
\end{equation}%
where $h_{i}$ is a bosonic \textquotedblleft holon\textquotedblright\
operator and $f_{i\sigma }$ a fermionic \textquotedblleft
spinon\textquotedblright\ operator. They satisfy the no-double-occupancy
constraint $\sum_{\sigma }f_{i\sigma }^{\dagger }f_{i\sigma }+h_{i}^{\dagger
}h_{i}=1$ at each lattice site.

The pseudogap phase in the RVB mean-field theory based on this slave-boson
formalism is characterized by a d-wave order parameter for the spinon
pairing \cite{sb2}

\begin{eqnarray}
\Delta ^{\mathrm{sb}} &=&\Delta _{x}^{\mathrm{sb}}=-\Delta _{y}^{\mathrm{sb}}
\notag \\
&=&\frac{3}{8}\left\langle f_{i\uparrow }f_{i+x\downarrow }-f_{i\downarrow
}f_{i+x\uparrow }\right\rangle  \label{sb2}
\end{eqnarray}%
which is finite at small doping below a characteristic temperature $T^{\ast
} $. The pseudogap phase is defined in the temperature regime of $T^{\ast
}>T>T_{c}$, where the superconducting transition temperature $T_{c}$ is
decided by the \textquotedblleft holon\textquotedblright\ condensation ($%
\left\langle h^{\dagger }\right\rangle \neq 0$) temperature in this
mean-field description. The corresponding phase diagram \cite{lee} is
schematically shown in Fig. 1.

Note that in the RVB mean-field theory \cite{sb1,sb2,lee}, besides the RVB
order parameter (\ref{sb2}), one also needs to introduce the following order
parameters $\kappa =\kappa _{x}=\kappa _{y}=\frac{3}{8}\left\langle
\sum_{\sigma }f_{i\sigma }^{\dagger }f_{i+x\sigma }\right\rangle $ and $%
B=\left\langle b_{i}b_{j}^{\dagger }\right\rangle =\left\langle
b_{i}^{\dagger }b_{j}\right\rangle .$ The detailed mean-field results
concerning the pseudogap phase will be presented in Sec. IV.

\begin{figure}[tp]
\begin{center}
\includegraphics{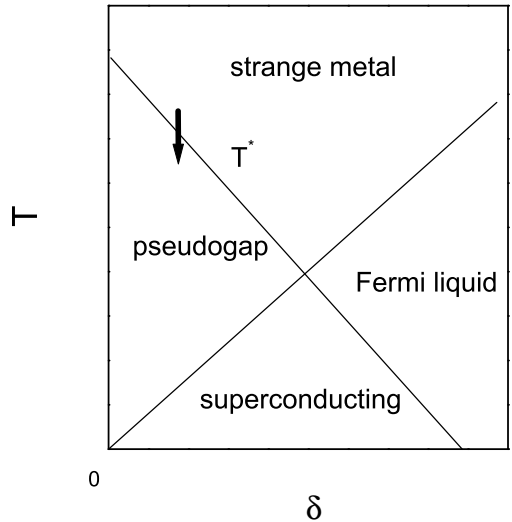} 
\end{center}
\caption{The phase diagram of the
slave-boson mean-field theory, which is divided into four regions by two
lines representing the RVB pairing ($\Delta ^{\mathrm{sb}}\neq 0$) and holon
condensation ($\left\langle h^{\dagger }\right\rangle \neq 0$),
respectively. The pseudogap phase is defined by the RVB pairing of spinons
below $T^{\ast },$ in the absence of the holon condensation. The later
determines the superconducting phase at low doping $\protect\delta $. See
Refs.\protect\cite{lee} for details. }
\label{fig1}
\end{figure}

\subsection{Pseudogap phase in the bosonic RVB state}

\subsubsection{Definition}

In the phase-string formalism of the $t-J$ model, the electron operator is
decomposed as \cite{string1} 
\begin{equation}
c_{i\sigma }=h_{i}^{\dagger }b_{i\sigma }e^{i\hat{\Theta}_{i\sigma }}.
\label{ps1}
\end{equation}%
In contrast to the slave-boson scheme (\ref{sb1}), both the
\textquotedblleft holon\textquotedblright\ $h_{i}$ and \textquotedblleft
spinon\textquotedblright\ $b_{i\sigma }$ here are \emph{bosonic} operators.
In such a \emph{bosonization} description, the fermionic commutation
relations of $c_{i\sigma }$'s can be restored by the phase-string factor $%
e^{i\hat{\Theta}_{i\sigma }}.$

As a distinctive feature of this decomposition, the definition of the
phase-string factor is given by $e^{i\hat{\Theta}_{i\sigma }}\equiv (-\sigma
)^{i}e^{i\frac{1}{2}\left[ \Phi _{i}^{b}-\sigma \Phi _{i}^{h}\right] }$,
where $\Phi _{i}^{b}=\Phi _{i}^{s}-\Phi _{i}^{0}$, with $\Phi
_{i}^{s}=\sum_{l\neq i}\theta _{i}\left( l\right) \sum_{\alpha }\alpha
n_{l\alpha }^{b}$ and $\Phi _{i}^{0}=\sum_{l\neq i}\theta _{i}\left(
l\right) $, while $\Phi _{i}^{h}=\sum_{l\neq i}\theta _{i}\left( l\right)
n_{l}^{h}.$ ($n_{l\alpha }^{b}$ and $n_{l}^{h}$ are spinon and holon number
operators. And $\theta _{i}\left( l\right) \equiv \func{Im}\ln \left(
z_{i}-z_{l}\right) $, with $z_{l}=x_{l}+iy_{l}.$) It can be verified that
the fermionic properties of $c_{i\sigma }$'s are ensured by such a phase
factor under the no-double-occupancy constraint $\sum_{\alpha }n_{l\alpha
}^{b}$ $+$ $n_{l}^{h}=1$.

The pseudogap phase in the bosonic-RVB (b-RVB) state \cite{string2} based on
the phase-string formalism is defined as follows. Firstly the d-wave
superconducting order parameter $\Delta _{\ }^{SC}$ can be written as \cite%
{string3} 
\begin{equation}
\Delta _{\ }^{SC}=\Delta ^{0}e^{i\Phi ^{s}}\   \label{ps2}
\end{equation}%
where the pairing amplitude

\begin{equation}
\Delta ^{0}=\left\langle h^{\dagger }\right\rangle ^{2}\Delta ^{s}
\label{ps3}
\end{equation}%
in which $\Delta ^{s}$ denotes the b-RVB order parameter, to be defined
below. Thus, \ compared to the slave-boson mean-field theory, the
superconducting order parameter has an extra phase factor $e^{i\Phi ^{s}}$
in Eq.(\ref{ps2}). When the holons experience a Bose condensation, with $%
\left\langle h^{\dagger }\right\rangle \neq 0$, the state is not necessarily
superconducting, if the phase $\Phi ^{s}$ still remains disordered.\cite%
{string3,string4}

Let us examine this from a \textquotedblleft top-down\textquotedblright\
approach. We note that the b-RVB order parameter $\Delta ^{s}$ will first
form at high temperature $T_{0}$ ($\sim J/k_{B}$ at small $\delta $).
Different from $\Delta ^{\mathrm{sb}}$ in the slave-boson theory,$\ \Delta
^{s}$ here no longer corresponds to the opening of an energy gap. It
characterizes short-range AF spin correlations according to $\left\langle 
\vec{S}_{i}\cdot \vec{S}_{j}\right\rangle _{\mathrm{NN}}=-3/8|\Delta
^{s}|^{2}$ and can persist over a wide range of temperatures and a finite
doping regime as illustrated in Fig. 2.

\begin{figure}[tp]
\begin{center}
\includegraphics{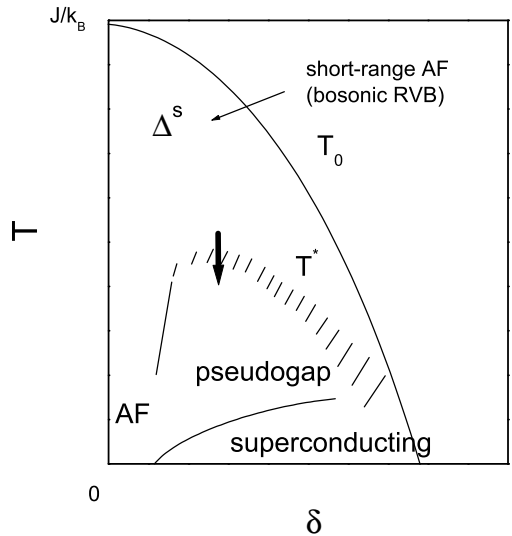} 
\end{center}
\caption{The phase diagram in the b-RVB
theory at low doping. From the top-down, the b-RVB pairing of spinons ($%
\Delta ^{s}\neq 0$) below $T_{0}$ characterizing the short-range AF
correlations. The pseudogap phase sets in below $T^{\ast }$ when the bosonic
holons experience the Bose condensation and thus the amplitude of the Cooper
pairing forms ($\Delta ^{0}\neq 0$). Eventually the superconducting phase is
realized when the phase coherence is achieved. Note that the phase coherence
factor in Eq.(\protect\ref{ps2}) is absent in the slave-boson RVB theory. }
\label{fig2}
\end{figure}
Then the pseudogap phase in this
framework is defined by \cite{string4} 
\begin{equation}
\Delta ^{0}\neq 0  \label{ps4}
\end{equation}%
The pairing amplitude $\Delta ^{0}$ becomes finite at a temperature $T^{\ast
}$ at which the bosonic holons are Bose condensed. On the other hand, the
superconducting transition occurs at $T_{c}$ when the phase coherence is
realized \cite{string3} in Eq.(\ref{ps2}), as the \textquotedblleft
spinon-vortices\textquotedblright\ in $\Phi ^{s}$ are bound together such
that $\Phi ^{s}$ is cancelled out at large-distance scales. Note that the
pseudogap phase between $T^{\ast }>T>T_{c}$ is also called \emph{spontaneous
vortex phase} in Ref. \cite{string4} since free spinon-vortices are unbound
in such a regime where the phase $\Phi ^{s}$ in Eq.(\ref{ps2}) is disordered
($T^{\ast }$ is denoted by $T_{v}$ in Ref. \cite{string4}). It has been
argued \cite{string4} that these free spinon-vortices will contribute to the
Nernst effect observed experimentally.

\subsubsection{Interplay between the kinetic and superexchange terms}

Different from the standard slave-boson mean-field theory, the spinon and
holon degrees of freedom in the b-RVB theory are intrinsically
\textquotedblleft entangled\textquotedblright\ together, which will be quite
essential in the following discussion of the properties of the pseudogap
phase. We shall first briefly discuss the basic theoretical structure of
this model below.

The low-energy effective Hamiltonian $H_{\mathrm{string}}=H_{h}+H_{s}$
derived \cite{string2} from the phase-string formalism of the $t-J$ model
(with $H_{h}$ and $H_{s}$ corresponding to the hopping term $H_{t}$ and the
superexchange term $H_{J}$, respectively) under the bosonic RVB order
parameter $\Delta ^{s}$ is given by 
\begin{equation}
H_{h}=-t_{h}\underset{\left\langle ij\right\rangle }{\sum }\left(
e^{iA_{ij}^{s}}\right) h_{i}^{\dagger }h_{j}+H.c.  \label{ps5}
\end{equation}%
\begin{equation}
H_{s}=-J_{s}\underset{\left\langle ij\right\rangle \sigma }{\sum }\left(
e^{i\sigma A_{ij}^{h}}\right) b_{i\sigma }^{\dagger }b_{j-\sigma }^{\dagger
}+H.c.  \label{ps6}
\end{equation}%
where $t_{h}\sim t$ and $J_{s}=J\Delta ^{s}/2\sim 0.5J,$ with $\Delta ^{s}=%
\underset{\sigma }{\sum }\left\langle e^{-i\sigma A_{ij}^{h}}b_{i\sigma
}b_{j-\sigma }\right\rangle _{NN}.$

An important feature of this effective model is a mutual entanglement
between the charge and spin degrees of freedom as mediated by the gauge
fields, $A_{ij}^{s}$ and $A_{ij}^{h},$ which satisfy the following
topological \textquotedblleft constraints\textquotedblright\ 
\begin{equation}
\sum_{c}A_{ij}^{s}=\pi \sum_{l\in c}\left( n_{l\uparrow }^{b}-n_{l\downarrow
}^{b}\right)  \label{ps8}
\end{equation}%
and 
\begin{equation}
\sum_{c}A_{ij}^{h}=\pi \sum_{l\in c}n_{l}^{h}\   \label{ps9}
\end{equation}%
for a closed loop $c$. Thus, a \textquotedblleft spinon\textquotedblright\
always sees a \textquotedblleft holon\textquotedblright\ as a $\pi $ fluxoid
and a \textquotedblleft holon\textquotedblright\ perceives a
\textquotedblleft spinon\textquotedblright\ as a $\pm \pi $ fluxoid ($\pm $
depends on the spin index).

In the pseudogap phase where the holons experience a Bose condensation, $%
A_{ij}^{h}$ can be reduced to a vector field describing a uniform flux
satisfying 
\begin{equation}
\sum_{\Box }A_{ij}^{h}\simeq \pi \delta .  \label{uflux}
\end{equation}%
Then $H_{s}$ in (\ref{ps6}) can be easily diagonalized.\cite{string2} In
such a case, a characteristic spinon gap (thus a spin gap) will be opened up
and the AF correlations will be suppressed at low energy (see Sec. IV).
Since the majority of spinons remain in RVB pairing at low temperatures, $%
A_{ij}^{s}$ will be substantially cancelled out in Eq.(\ref{ps8}), which in
turn is in favor of the holon condensation in $H_{h}$ \emph{in a
self-consistent way}. So in the pseudogap phase, \emph{the kinetic energy of
the holons is expected to get improved at the expense of the AF correlations}
according to the phase-string description.

On the other hand, in the \textquotedblleft normal\textquotedblright\ state
where the holon condensation is gone, the spinon gap will disappear and
there will be \emph{strong fluctuations} in $A_{ij}^{s}$ due to unpaired $%
\pm \pi $ fluxoid bound to \emph{thermally} \emph{excited} spinons in terms
of Eq.(\ref{ps8}). This will then lead to a great deal of frustrations on
the holon part through $H_{h},$ again self-consistently, causing\emph{\
holons lose phase coherence and behave incoherently}. In contrast to such a
frustration of the kinetic energy of the charge carriers, strong AF
correlations will be \emph{restored} in $H_{s}\,$instead$,$ where the gauge
field $A_{ij}^{h}$ can no longer treated as in Eq.(\ref{uflux}), but rather
as describing $\pi $ fluxoids bound to randomly distributed holons according
to Eq.(\ref{ps9}), whose frustration effects on the AF correlations are
found to be \emph{minimal}, in contrast to the case of Eq.(\ref{uflux}) in
the pseudogap phase.

Therefore, one will find a distinctive interplay between the charge and spin
degrees of freedom in the pseudogap state and \textquotedblleft
normal\textquotedblright\ state based on the framework of the b-RVB theory,
characterized by Eqs.(\ref{ps5}) and (\ref{ps6}). Namely, the AF
correlations and the kinetic energy are mutually repulsive to each other,
whose competition becomes the driving force for the formation of the
pseudogap phase at low temperatures. The detailed results will be presented
in the following section.

\section{Results}

\subsection{Exchange-energy-driven vs. kinetic-energy-driven}

In the above we have introduced the definitions of pseudogap state in two
different mean-field approaches in the $t-J$ model. In this section, we
explore and compare the driving mechanisms for such pseudogap states in the
slave-boson and b-RVB descriptions, respectively, and show that two
mechanisms are drastically distinct. Namely, one is exchange-energe-driven
and the other is kinetic-energy-driven.

\subsubsection{Superexchange energy}

Let us consider the superexchange energy and examine the change of the
superexchange energy during the formation of a pseudogap state according to
the mean-field theories outlined in Sec. III.

Firstly let us focus on the pseudogap state in the slave-boson RVB theory,
which is featured by the opening of the slave-boson RVB gap $\Delta ^{%
\mathrm{sb}}.$ Since the origin of $\Delta ^{\mathrm{sb}}$ comes entirely
from $H_{J}$, the superexchange energy is expected to be gained in the
pseudogap phase.

The calculation based on the mean-field theory outlined in Sec. III is
standard and straightforward \cite{sb2}. Fig. 3(a) shows the results
calculated at $\delta =0.125$ in a $32\times 32$ square lattice. Clearly the
superexchange energy is reduced with $\Delta ^{\mathrm{sb}}\neq 0$ (solid
curve) as compared to the state with $\Delta ^{\mathrm{sb}}=0$ (dotted
curve) below the onset temperature $T^{\ast }$.

\begin{figure}[tbp]
\begin{center}
\includegraphics{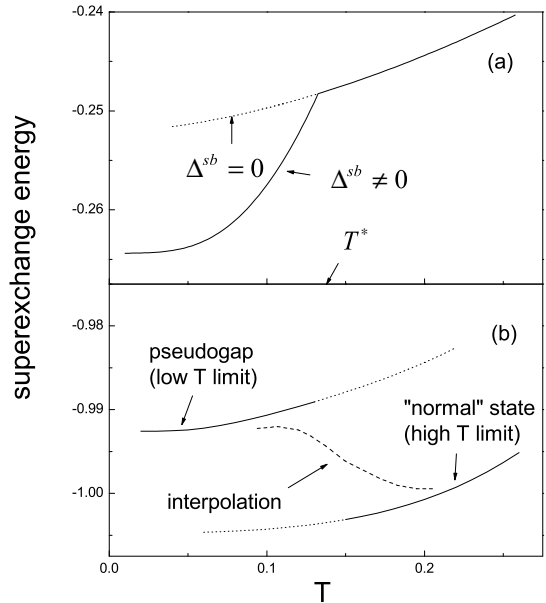} 
\end{center}
\caption{The superexchange energies calculated in the slave-boson
mean-field theory (a) and in the b-RVB theory (b). Two results show the 
\emph{opposite} trends, indicating that while it is
superexchange-energy-driven in the former, it is not in the latter. Note
that the temperature in the horizontal axis is plotted in units of $J$ in
(a) and $2J_{s}(\sim J)$ in (b).}
\label{fig3}
\end{figure}

The result that the pseudogap phase is superexchange-energy-driven in the
slave-boson theory is within the expectation as mathematically the fermionic
spinon part closely resembles a BCS mean-field state where the attractive
potential is obviously the driving force. Such a fact should remain true
even if one includes the gauge fluctuations in both \textrm{U(1) }and 
\textrm{SU(2)} formalisms\cite{lee,su(2)}. Usually the gauge fluctuations
will get suppressed in the pseudogap phase with the opening of the gap $%
\Delta ^{\mathrm{sb}}.$ Such a competition between the pseudogap $\Delta ^{%
\mathrm{sb}}$ and gauge fluctuations may quantitatively modify the
superexchange energy, but not the driving mechanism itself. \textrm{\ }

Now let us consider the b-RVB state, in which the pseudogap phase is marked
by the occurrence of the holon condensation. As discussed in Sec. III B, the
corresponding effect of the holon condensation on the superexchange energy
will be mediated through the topological gauge field $A_{ij}^{h}$ in Eq.(\ref%
{ps6}). Let us consider two extreme limits. In one limit, let all holons be
condensed such that $A_{ij}^{h}$ describes a uniform flux [Eq.(\ref{uflux}%
)], which is deep in the pseudogap regime. In the opposite limit, all holons
behave incoherently such that $A_{ij}^{h}$ describes a \emph{randomly
distributed} $\pi $ fluxoids on the lattice (of a total number $N\delta $),
which approximately represents the case of the high-temperature
\textquotedblleft normal\textquotedblright\ state. Fig. 3(b) illustrates the
superexchange energy determined by the mean-field solution of $H_{s}$ as a
function of $T,$ with $A_{ij}^{h}$ being treated in the above-mentioned two
limits, in a $32\times 32$ lattice. As shown by Fig. 3(b), it will always 
\emph{cost} the superexchange energy going into the pseudogap state from the
high-$T$ \textquotedblleft normal\textquotedblright\ state.

Note that in the temperature range between these two limits, $A_{ij}^{h}$
itself should behave more complicated than the above simple-minded
treatments, which in general involves the holon dynamics. It is usually
difficult to determine the true superexchange energy as a function of
temperature in a self-consistent way, whose realistic behavior is sketched
by a long dashed curve in Fig. 3(b). Nevertheless, in the two opposite
limits at low- and high-temperatures the results present in Fig. 3(b) should
be reliable, which unambiguously indicate that the pseudogap state is formed
at the expense of the superexchange energy in the b-RVB theory, in contrast
to the superexchange-energy-driven picture for the slave-boson mean-field
theory shown in Fig. 3(a).

\subsubsection{Kinetic energies}

Consistent with the divergent behaviors of the superexchange energies in two
theories, the kinetic energies for the two RVB mean-field states also show
opposite trends upon entering the pseudogap phases.
\begin{figure}[tbp]
\begin{center}
\includegraphics{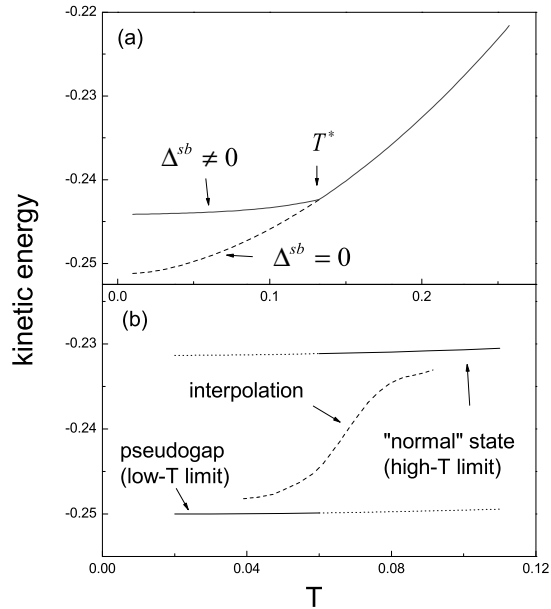} 
\end{center}
\caption{The kinetic energies in the slave-boson mean-field
theory (a) and b-RVB theory (b). The results are consistent with the
superexchange energies given in Fig. 3, showing that the pseudogap phase is
kinetic-energy-driven in the b-RVB theory, whereas it is
superexchange-energy-driven in the slave-boson mean-field theory. Note that $%
T$ axis is in units of $J$ in (a) while in units of $t_{h}$ in (b).}
\label{fig4}
\end{figure}

First of all, the kinetic energy in the slave-boson mean-field state can be
also straightforwardly calculated as shown in Fig. 4(a). Obviously the
kinetic energy is lost with the opening the RVB gap $\Delta ^{\mathrm{sb}}$.
Thus, in combination with Fig. 3(a), it is established that the pseudogap
phase in the slave-boson RVB theory is indeed \emph{%
superexchange-energy-driven.}

Next, we consider the kinetic energy based on $H_{h}$ in the b-RVB theory.
Here $H_{h}$ describes that free bosonic holons hop in the presence of a
gauge field $A_{ij}^{s}$, which depicts $\pm \pi $ flux-tubes bound to
spinons. At low temperatures, when spinons are well paired in the RVB state, 
$A_{ij}^{s}$ will get suppressed. In this case, the bosonic holons will
experience a Bose condensation. This is the pseudogap phase defined in the
b-RVB state as discussed in Sec. III B. Here a spin gap is opened up in the
spin excitation spectrum such that to break up an RVB pair or equivalently a
vortex-anti-vortex bound pair in $A_{ij}^{s}$ will cost an energy $E_{g}$
(see Fig. 8 below)$.$ It means that the fluctuations in $A_{ij}^{s}$ will be
suppressed at low temperatures, which then enforces the holon condensation
in $H_{h}$ in a self-consistent fashion. In the opposite limit, when a lot
of free spinon excitations are created at high temperatures, strong
fluctuations of $A_{ij}^{s}$ due to the free $\pm \pi $ flux-tubes bound to
excited spinons will destroy the bosonic holons' phase coherence and cause
them behave incoherent.
\begin{figure}[tbp]
\begin{center}
\includegraphics{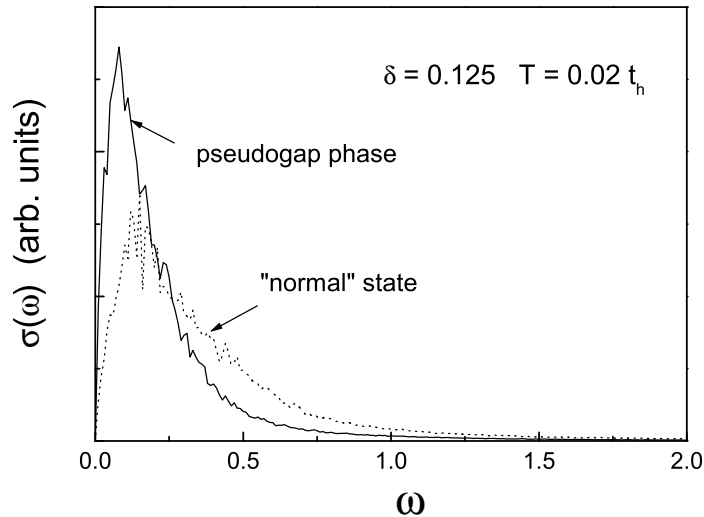} 
\end{center}
\caption{The
optical conductivity in the b-RVB theory calculated in the pseudogap phase
and \textquotedblleft normal\textquotedblright\ phase as defined in the
text. The spectral weight shifts towards the low energies as it enters the
pseudogap phase. Here the $\protect\omega $-axis is ploted in units of $t_{h}
$.}
\label{fig5}
\end{figure}

Thus, we can calculate the kinetic energy based on $H_{h}$ in such two
extreme limits, similar to the preceding discussion of the superexchange
term. In the deep pseudogap limit, one may simply treat $A_{ij}^{s}=0$ and
the holons as non-interacting bosons$.$ In the high-temperature
\textquotedblleft normal\textquotedblright\ state, one can treat $A_{ij}^{s}$
as describing $\pm \pi $ flux-tubes bound to free spinons, which are
randomly distributed on lattice to simulate those excited from the RVB
background with a number $n_{\mathrm{spinon}}^{\mathrm{free}}\left( T\right) 
$. Generally $n_{\mathrm{spinon}}^{\mathrm{free}}\left( T\right) $ should
increase with the temperature. For our purpose, we can define a prototype
\textquotedblleft normal\textquotedblright\ state with $n_{\mathrm{spinon}}^{%
\mathrm{free}}$ fixed at $2\delta N$ (note that $n_{\mathrm{spinon}}^{%
\mathrm{free}}\sim \delta N$ at $T=T_{c}$ \cite{string2,string3}), and
calculate the average kinetic energy $\left\langle H_{h}\right\rangle $ as a
function of temperature.

Fig. 4(b) shows the results. The kinetic energy $\left\langle
H_{h}\right\rangle $\ does decrease with the decrease of $n_{\mathrm{spinon}%
}^{\mathrm{free}},$ which means that with the decrease of the temperature,
the kinetic energy of holons will get continuously improved due to the
suppression of the fluctuations in $A_{ij}^{s}.$ Again the realistic case
will be some kind of interpolation between the solid curves shown in Fig.
4(b). But it does not change the general trend that the kinetic energy will
be gained by entering the pseudogap phase, \emph{i.e.}, a
kinetic-energy-driven mechanism, which is consistent with the loss of the
superexchange energy discussed earlier on.

Finally, we have computed the optical conductivity in the b-RVB theory.
Based on the same approximation in dealing with $H_{h}$ above, the optical
conductivity is obtained in Fig. 5 for two cases of the pseudogap and
\textquotedblleft normal\textquotedblright\ states with $n_{\mathrm{spinon}%
}^{\mathrm{free}}\sim \delta N$ and $n_{\mathrm{spinon}}^{\mathrm{free}}\sim
2\delta N,$ respectively. It shows that the low-energy spectral weight does
increase in the pseudogap phase as compared to the normal state, in
consistency with the kinetic energy behavior.

\subsection{\protect\bigskip Spin-spin correlations}

The sharp contrast between the qualitative behaviors of the superexchange
and kinetic energies in two mean-field theories is quite remarkable, as they
are supposed to describe the same $t-J$ model. In the following, we analyze
the detailed spin-spin correlations in order to understand the physics
underlying such distinctions.

\subsubsection{Uniform spin susceptibility}

The uniform spin susceptibility $\chi _{u}$ in the cuprates has generally
exhibited a suppression at low temperatures in the underdoped phase. This
behavior has served as one \cite{millis} of the main experimental evidence
for the existence of a pseudgap above $T_{c}$.

In the slave-boson RVB theory, the opening of the RVB gap $\Delta ^{\mathrm{%
sb}}$ will naturally result in the suppression of $\chi _{u}$ at low
temperature as shown in Fig. 6(a). Similarly, in the b-RVB state, the
uniform spin susceptibility $\chi _{u}$ also shows a continuous reduction at
low temperatures, as given in Fig. 6(b), where the curves calculated in the
two limits of the pseudogap and \textquotedblleft normal\textquotedblright\
states (corresponding to Fig. 3(b)) are presented for comparison.

Here we see that the pseudogap behavior in the uniform spin susceptibility
is not very sensitive to the detailed mechanisms and the general trend
remains essentially the same. This implies that the density of spin states
near momentum $(0,0)$ is indeed universally suppressed in both cases.
However, the situation will become quite different when we consider
spin-spin correlations near the AF momentum $(\pi ,\pi )$ below$.$\bigskip 
\begin{figure}[tbp]
\begin{center}
\includegraphics{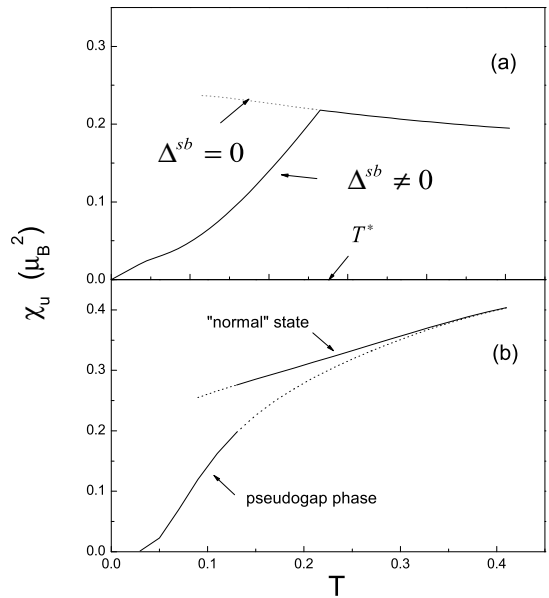} 
\end{center}
\caption{A comparison of the uniform spin
susceptibilities in the slave-boson mean-field theory (a) and b-RVB theory
(b). Both show a suppression at low temperatures, indicating the openning up
of a spin gap near momentum $(0,0)$. The $T$-axis is scaled with $J$ in (a)
and $2J_{s}$ in (b).}
\label{fig6}
\end{figure}

\subsubsection{Equal-time spin-spin correlations}

The superexchange energy is related to the equal-time NN spin-spin
correlations in the $t-J$ model. In the salve-boson RVB mean-field state,
the equal-time the spin-spin correlation function is shown in Fig. 7(a). An
enhancement of the NN spin correlations is the direct reason that the
superexchange energy is gained in the pseudogap state. It is noted that the
overall AF correlations are quite weak for both the normal and pseudogap
states in Fig. 7(a), which can be visibly strengthened after the Gutzwiller
projection is applied.

Similarly, one can compute the equal-time spin correlation function in the
b-RVB description. The results for the pseudogap and normal states are
presented in Fig. 7(b), respectively, in the two limits shown in Fig. 3(b),
indicates that the spin-spin correlations are clearly suppressed in the
pseudogap state, which is responsible for the increase of the superexchange
energy in Fig. 3(b).
\begin{figure}[tbp]
\begin{center}
\includegraphics{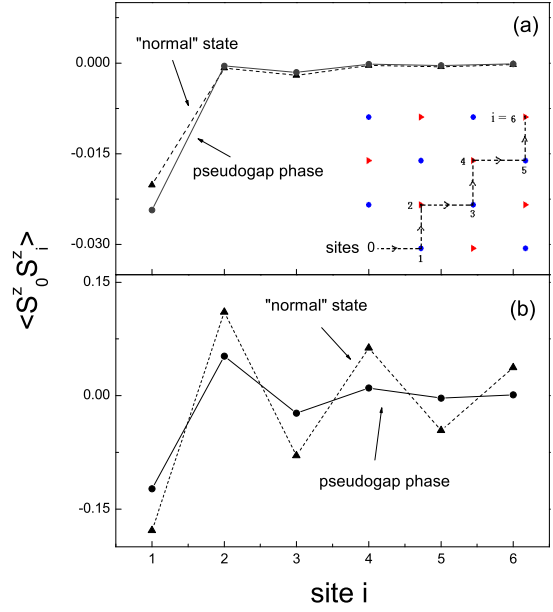} 
\end{center}
\caption{Equal-time
spin-spin correlation function in the slave-boson theory (a) and b-RVB
theory (b) in both the pseudogap and "normal" states. All results here are
compared at $T=0$.}
\label{fig7}
\end{figure}

At this step, we can see that the main distinction between two pseudogap
states is closely related to the AF correlations. In the slave-boson
mean-field state, the short-range AF correlations are rather weak in the
normal state, and by opening an RVB gap, the spin-spin correlations near $%
\left( \pi ,\pi \right) $ will get improved (note that the comparison is
made at the \emph{same} temperature by extrapolation to $T=0$). This general
trend will not be altered even if one introduces the Gutzwiller projection,
although the overall AF correlations can be enhanced by the projection for 
\emph{both} cases. In contrast, the AF correlations in the normal phase of
the b-RVB state are already quite strong, and are always suppressed by
entering the pseudogap phase, thus leading to the superexchange-energy loss.

So the change of the AF correlations will hold the key to understanding the
mechanisms for the above pseudogap phases. Generally speaking, weak AF
correlations in the spin background favor the hopping of the holes. This is
the reason why the kinetic energy in the slave-boson mean-field state is
better in the normal state, whereas is reduced as the RVB gap $\Delta ^{%
\mathrm{sb}}$ opens up with the improved short-range AF correlations. On the
other hand, the AF correlations are much stronger in the normal phase of the
b-RVB state, which is not in favor of the kinetic energy. So the holons can
only hop incoherently under the influence of the randomly distributed $\pm
\pi $ fluxoids via $A^{s}$ in the phase-string description. In the pseudogap
phase of the b-RVB state, the hopping of the holons becomes coherent and the
holons are condensed with better kinetic energies, at the expenses of the AF
correlations which become gapped (Fig. 8 below).

In the following, the change in the AF oscillations of $\left\langle
S_{i}^{z}S_{j}^{z}\right\rangle $ can be further analyzed in terms of the
weight transfer in the dynamic susceptibility function $\func{Im}\chi
^{zz}\left( \vec{q},\omega \right) $ at different energies for a fixed $\vec{%
Q}_{\mathrm{AF}}=\left( \pi ,\pi \right) ,$ according to the Fourier
transformation of Eq.(\ref{sumrule2}).

\subsubsection{Dynamic spin susceptibility near $\left( \protect\pi ,\protect%
\pi \right) $}

In the pseudogap phase based on the bosonic RVB description, the equal-time
AF correlations have been shown to be suppressed. In the following, we
further investigate how such a suppression is exhibited as a function of the
frequency $\omega $ near the AF momentum $\left( \pi ,\pi \right) .$

Fig. 8 shows $\func{Im}\chi ^{zz}\left( \vec{q},\omega \right) $ at $\vec{Q}%
_{\mathrm{AF}}=\left( \pi ,\pi \right) $ in the two extreme limits of the
pseudogap and \textquotedblleft normal\textquotedblright\ states shown in
Fig. 3(b). Note that the temperature is set at zero for both cases, just for
convenience. In the pseudogap phase, a resonance-peak at $\vec{Q}_{\mathrm{AF%
}}$ emerges around $E_{g}\sim 0.5(2J_{s})$ at $\delta =0.125.$ In the
previous work, \cite{string2} such a sharp-peak structure has been used to
explain the so-called $41$ $\mathrm{meV}$ resonance-like peak observed by
the inelastic neutron-scattering in the optimal-doped \textrm{YBCO }compound 
\cite{neutron1}, where $T^{\ast }$ seems to coincide with $T_{c}.$ A broader
peak feature has been also found \cite{neutron2} in the underdoped \textrm{%
YBCO }compounds at $T_{c}<T<T^{\ast },$ and the present theory provides a
natural description. Note that the sharpness of the peak in Fig. 8 is due to
the artifact in treating the gauge field $A_{ij}^{h}$ by assuming the holons
are all in an ideal Bose condensation. It can be easily broadened once the
density fluctuations of holons are considered.

The resonance-like peak around $E_{g}$ disappears in Fig. 8 in the
\textquotedblleft normal\textquotedblright\ state when the holon coherence
is gone and the holons behave like incoherent objects, such that $A_{ij}^{h}$
can be treated as describing randomly distributed $\pi $ flux-tubes as
discussed in Sec. III B. The dashed curve show the corresponding result
artificially extrapolated to $T=0,$ with the remaining spectral weight at $%
\vec{Q}_{\mathrm{AF}}$ shifting down to the low frequency $\omega \sim 0.$
This is consistent with the fact that the AF correlations will be enhanced
in the \textquotedblleft normal\textquotedblright\ state.

\begin{figure}[tbp]
\begin{center}
\includegraphics{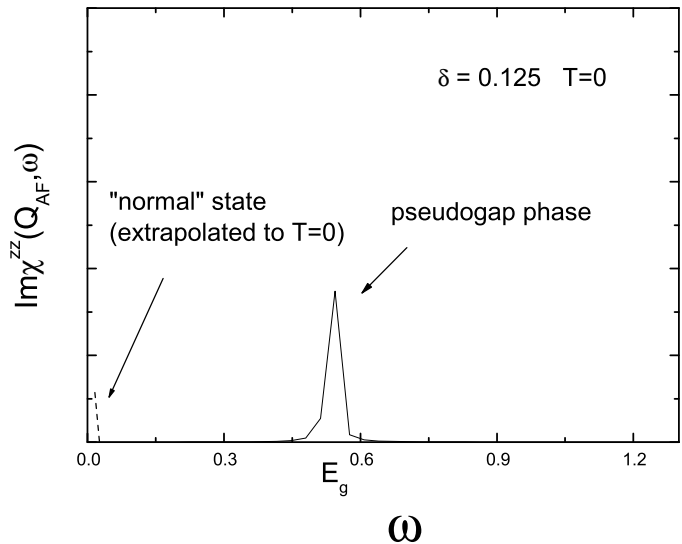} 
\end{center}
\caption{Dynamic spin susceptibility
obtained in the b-RVB theory at $Q_{AF}=(\protect\pi ,\protect\pi )$. A
resonance-like peak is found an energy $E_{g}\sim \protect\delta J$ in the
pseudogap phase, which disappears in the high-$T$ \textquotedblleft
normal\textquotedblright\ state where the weight in $\func{Im}\protect\chi (%
\protect\omega )$ is shifted to $\protect\omega \sim 0$ when is extrapolated
to $T=0.$ The $\protect\omega $-axis is in units of $2J_{s}\sim J$. }
\label{fig8}
\end{figure}

We can further study the AF correlations near $\omega \sim 0$ by calculating 
$\func{Im}\chi ^{zz}\left( \vec{q},\omega \right) /\omega $ at $\vec{Q}_{%
\mathrm{AF}}$ and $\omega \rightarrow 0.$ Note that such a quantity is
related to the NMR spin relaxation rate $1/T_{1}T$ for $^{63}$\textrm{Cu }%
nuclear spins (a $\vec{q}$-dependent hyperfine coupling factor is neglected
here for simplicity). The results for the two extreme limits considered
before are shown in Fig. 9: the main panel is for the \textquotedblleft
normal\textquotedblright\ state\ case, which demonstrates a non-Korringa
behavior often observed \cite{NMR} in the cuprates due to strong AF
correlations near $\vec{Q}_{\mathrm{AF}}$ and $\omega \sim 0.$ By contrast,
a spin gap behavior of $1/T_{1}T$ also seen in experiment \cite{millis} is
indeed exhibited in the pseudogap state as shown in the inset of Fig. 9.
Both are qualitatively consistent with the experimental measurements in the
cuprate superconductors.
\begin{figure}[tp]
\begin{center}
\includegraphics{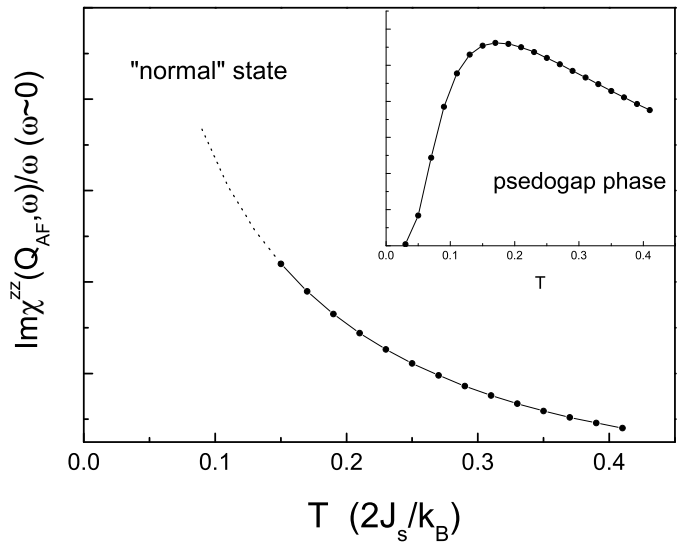} 
\end{center}
\caption{$\func{Im%
}\protect\chi (\vec{Q}_{\mathrm{AF}},\protect\omega )/\protect\omega $ at $%
\protect\omega \rightarrow 0$ is shown as a function of $T$ in the b-RVB
theory$.$ Such a quantity is approximately related to the spin relaxation
rate in NMR measurement. }
\label{fig9}
\end{figure}

So far we have only considered the dynamic spin susceptibility function
based on the b-RVB theory. Note that the AF correlations are generally quite
weak in the slave-boson mean-field theory [see Fig. 7(a)] and there are not
much features near $\vec{Q}_{\mathrm{AF}}$ to give rise to the
experimentally interested properties like $1/T_{1}T,$ etc, at the mean-field
level. One has really go beyond the mean-field approximation here. For
example, a resonance-like peak structure around $\vec{Q}_{\mathrm{AF}}$ has
been obtained in the superconducting phase after a modified
random-phase-approximation approach \cite{brinkman}. This is beyond the
scope of the present work. Nevertheless, the conclusion that the
superexchange energy will be \emph{gained} by the opening of the pseudogap $%
\Delta ^{\mathrm{sb}}$ should still hold true, as discussed in Sec. IV A, as
responsible by an enhancement of the equal-time (frequency-integrated) spin
correlations at a large momentum around $\left( \pi ,\pi \right) $ [opposed
to the suppression of the uniform susceptibility near ($0,0)]$.

\section{Discussion and conclusions}

In this paper, we have shown that the recent optical measurements of the
cuprates have clearly demonstrated that both the pseudogap and
superconducting phases are driven by the kinetic energy within the $t-J$
model description, based on a general consideration of the optical sum rule
in the Hubbard model and its relation to the $t-J$ model. Under such an
analysis, by entering the pseudogap phase, the superexchange energy of the $%
t-J$ model should get suppressed instead.

Then, we have examined two pseudogap phases obtained in the slave-boson RVB
and b-RVB mean-field states of the $t-J$ model. Although the uniform spin
susceptibility is universally suppressed in both cases, which is consistent
with experiment, the underlying driving mechanisms for the pseudogap phases
are found to be precisely opposite. Namely, it is
superexchange-energy-driven in the slave-boson RVB theory as opposed to the
kinetic-energy-driven in the b-RVB theory. If the above general
consideration based on the optical sum rules is correct, then we conclude
that the superexchange-energy-driven mechanism in the slave-boson RVB state
is not supported experimentally.

The distinction between the two cases is very basic and physically
revealing. It can be traced back to the central issue concerning the AF
correlations. In the high-$T$ phase (\textquotedblleft
normal\textquotedblright\ state) of the slave-boson mean-field state, the AF
correlations are relatively quite weak, which is fairly favorable to the
kinetic energy of holes as the latter can easily hop on the uniform RVB
background. In the low-$T$ phase (pseudogap state), an RVB pairing of
spinons emerges, which improves the superexchange energy as well as the
short-range AF correlations (or more precisely, the spin correlations at
large-momenta), whereas the holes become less easy to hop around with the
increasing kinetic energy.

By contrast, there already exist strong AF correlations in the high-$T$
phase (``normal'' state) of the b-RVB state, characterized by the b-RVB
order parameter $\Delta ^{s}$ which can persist up to $T\sim J/k_{B}\sim
1,500$ $\mathrm{K}$ at low doping. In this regime, the kinetic energy of
holes is actually strongly suppressed and the holons behave like incoherent
objects moving in a short-range AF correlated spin background. This is quite
different from the above slave-boson RVB state. Only in the low-$T$
pseudogap phase can the phase coherence of the holons be restored with
gaining the kinetic energy. It occurs at the expense of the low-energy AF
correlations, or the superexchange energy, and thus is entirely
kinetic-energy-driven.

Mathematically, the whole interplay is straightforward in the slave-boson
mean-field theory. However, the interplay in the b-RVB theory is different:
due to the presence of strong AF correlations, the low-energy effective
theory is no longer a mean-field one by nature. The charge and spin degrees
of freedom are generally \emph{entangled} together by a mutual
Chern-Simons-type gauge structure, reflecting the phase string effect as
described in Eqs.(\ref{ps5}) and (\ref{ps6}). Thus the pseudogap and
\textquotedblleft normal\textquotedblright\ states differ not simply by some
mean-field order parameters, as in the slave-boson RVB states.

Here strong AF correlations will influence the hopping of holons through the
topological gauge field $A_{ij}^{s},$ as if each excited spinon is a $\pi $
fluxoid, which severely frustrate the phase coherence of the holons. On the
other hand, when the bosonic holons recover their phase coherence and are
Bose condensed at low-$T$, they will affect the spinon degrees of freedom
drastically through the topological gauge field $A_{ij}^{h}$ in Eq.(\ref{ps6}%
), going from a random distributed $\pi $ fluxoids bound to the incoherent
holons at high-$T$ to the uniform flux given in Eq.(\ref{uflux}) in the
holon condensed (pseudogap) phase. The AF correlations at low-energy is
subsequently suppressed (gapped) in the pseudgap phase relative to the high-$%
T$ phase, pushing the weight up towards a higher energy $E_{g}$ to form a
\textquotedblleft resonance-like\textquotedblright\ peak as shown in Fig. 8
and to result in the reduction of the NMR spin relaxation rate in Fig.9.
Self-consistently, the opening of the spinon gap at low-energies will
substantially reduce the fluctuations of $A_{ij}^{s}$ in the hopping term
and improve the kinetic energy as shown in Fig. 4(b) as well as the
low-energy optical conductivity in Fig. 5, which further strengths the holon
Bose condensation, and thus the pseudgap phase.

\begin{acknowledgments}
We acknowledge helpful discussions with W. Q. Chen and X. L. Qi. This work
is supported by the grants of NSFC, the grant no. 104008 and SRFDP from MOE
of China.
\end{acknowledgments}


\begin{thebibliography}{99}
\bibitem{timusk} For a review, see, T. Timusk and B. Statt, Rep. Prog. Phys. 
\textbf{62}, 61 (1999).

\bibitem{f-RVB1} P. W. Anderson, Science \textbf{235}, 1196 (1987).

\bibitem{f-RVB2} G. Baskaran, Z. Zou, P. W. Anderson, Solid State Comm. 
\textbf{63}, 973 (1987).

\bibitem{f-RVB3} P. W. Anderson, P. A. Lee, M. Randeria, T. M. Rice, N.
Trivedi, F. C. Zhang, cond-mat/0311467.

\bibitem{lps1} V. J. Emery and S. A. Kilvelson, Nature \textbf{374}, 434
(1995).

\bibitem{lps2} J. Orenstein and A. J. Millis, Science \textbf{288}, 468
(2000).

\bibitem{preform} S. Alexandrov and N. F. Mott, \emph{High Temperature
Suerpconductors and Other Superfluids }(Talor and Francis, London, 1994).

\bibitem{ddw} S. Chakravarty, R. B. Laughlin, D. K. Morr and C. Nayak, Phys.
Rev. B \textbf{63}, 094503 (2001).

\bibitem{super1} P. W. Anderson, Science \textbf{279}, 1196 (1998); Physica
C \textbf{341-348}, 9 (2000).

\bibitem{super2} J. E. Hirsch, Physica C \textbf{199}, 305 (1992).

\bibitem{so51} D. J. Scalapino and S. R. White, Phys. Rev. B \textbf{58},
8222 (1998).

\bibitem{so52} E. Demler and S. C. Zhang, Nature \textbf{396}, 733 (1998).

\bibitem{super3} J. E. Hirsch, Science \textbf{295}, 2226\textbf{\ }(2002).

\bibitem{super4} M. R. Norman and C. P\'{e}pin, Phys. Rev. B \textbf{66},
100506 (2002)

\bibitem{super5} T. K. Kope\'{c}, Phys. Rev. B \textbf{67}, 014520 (2003).

\bibitem{super6} S. Chakravarty, H.Y. Kee and E. Abrahams, Phys. Rev. B 
\textbf{67}, 100504(R) (2003).

\bibitem{super7} C. C. Homes and S. V. Dordevic, D.A. Bonn, R. Liang, W.N.
Hardy, Phys. Rev. B \textbf{69}, 024514 (2004).

\bibitem{sumrule} J. E. Hirsch and F. Marciglio, Phys. Rev. B \textbf{62},
15131 (2000).

\bibitem{exp} H. J. A. Molegraaf, C. Presura, D. van der Marel, P. H. Kes,
and M. Li, Science \textbf{295},,2239 (2002).

\bibitem{sb1} Z. Zou and P. W. Anderson, Phys. Rev. B \textbf{37,} 627 (1988)

\bibitem{sb2} G. Kotliar and J. Liu, Phys. Rev. B \textbf{38}, 5142 (1988)

\bibitem{string1} Z.Y. Weng, D.N. Sheng, Y.C. Chen, and C. S. Ting, Phys.
Rev. B \textbf{55}, 3894 (1997).

\bibitem{string2} Z. Y. Weng, D. N. Sheng and C. S. Ting, Phys. Rev. Lett. 
\textbf{80}, 5401 (1998); Phys. Rev. B \textbf{59}, 8943 (1999).

\bibitem{neutron1} H. F. Fong, B. Keimer, P. W. Anderson, D. Reznik, F.
Dogan and I. A. Aksay, Phys. Rev. Lett. \textbf{75}, 316 (1995).

\bibitem{neutron2} H. F. Hong, P. Bourges, Y. Sidis, L. P. Regnault, J.
Bossy, A. Ivanov, D. L. Milius, I. A. Aksay and B. Keimer, Phys. Rev. B 
\textbf{61}, 14773 (2000).

\bibitem{so53} P. Dai, H. A. Mook, S. M. Hayden, G. Aeppli, T. G. Perring,
R. D. Hunt, and F. Doan, Science \textbf{284}, 1344 (1999).

\bibitem{bza} G. Baskaran, Z. Zou, and P. W. Anderson, Solid State Commun. 
\textbf{63}, 973 (1987).

\bibitem{string4} Z. Y. Weng and V. N. Muthukumar, Phys. Rev. B \textbf{66},
094509 (2002).

\bibitem{pseudo2} W. Rantner and X. G. Wen, Phys. Rev. B \textbf{66}, 144501
(2002).

\bibitem{c-axis1} P. W. Anderson, \emph{The Theory of Superconductivity in
the High }$T_{c}$\emph{\ Cuprates}, (Princeton Univ. Press, Princeton, 1997).

\bibitem{c-axis2} S. Chakravarty, H. Y. Kee, and E. Abrahams, Phys. Rev.
Lett. \textbf{82}, 2366 (1999).

\bibitem{string3} V. N. Muthukumar and Z. Y. Weng, Phys. Rev. B \textbf{65},
174511 (2002).

\bibitem{lee} P. A. Lee and N. Nagaosa, Phys. Rev. B \textbf{46}, 5621
(1992).

\bibitem{su(2)} X. G. Wen and P. A. Lee, Phys. Rev. Lett. \textbf{76}, 503
(1996); P.A. Lee, N. Nagaosa, T.K. Ng, and X.-G. Wen, Phys. Rev. B \textbf{57%
}, 6003 (1998).

\bibitem{millis} A. J. Millis and H. Monien, Phys. Rev. Lett. \textbf{70},
2810 (1993).

\bibitem{NMR} R. E. Walstedt, B. S. Shastry, and S-W. Cheong, Phys. Rev.
Lett. \textbf{72}, 3610 (1994), and references therein.

\bibitem{brinkman} J. Brinckmann and P. A. Lee, Phys. Rev. Lett. \textbf{82}%
, 2915 (1999).
\end{thebibliography}
\end{document}